\documentclass[aps,pre,groupedaddress,amsmath,twocolumn,eqsecnum,showpacs]{revtex4}
   \usepackage{bm}
\usepackage{graphicx,amssymb}
\usepackage{bm}
\usepackage{amsmath}
\def\beq{\begin{equation}}
\def\eeq{\end{equation}}
\def\beqa{\begin{eqnarray}}
\def\eeqa{\end{eqnarray}}
\def\zero{{(0)}}
\def\one{{(1)}}
\def\two{{(2)}}

\def\N{\Gamma}
\newcommand{\nn}{\nonumber\\}
\newcommand{\bb}{\begin{equation}}
\newcommand{\ee}{\end{equation}}
\newcommand{\ba}{\begin{eqnarray}}
\newcommand{\ea}{\end{eqnarray}}
\newcommand{\dd}{{\rm d}}

\newcommand{\rr}{{\mathbf r}}

\newcommand{\mo}{\langle\sigma\rangle}
\newcommand{\mt}{\langle\sigma^2\rangle}
\newcommand{\mth}{\langle\sigma^3\rangle}
\newcommand{\mn}{\langle\sigma^n\rangle}

\begin{document}
%\draft
\title{Multicomponent fluid of hard spheres near a wall}
\author{Alexandr Malijevsk\'y}
\email{malijevsky@icpf.cas.cz} \affiliation{E. H\'ala Laboratory of
Thermodynamics, Academy of Science of the Czech Republic, Prague 6,
Czech Republic \\ Institute of Theoretical Physics, Faculty of
Mathematics and Physics, Charles University, Prague 8, Czech
Republic}
\author{Santos B. Yuste}
\email{santos@unex.es} \homepage{http://www.unex.es/fisteor/santos/}
\author{Andr\'es Santos}
\email{andres@unex.es} \homepage{http://www.unex.es/fisteor/andres/}
\affiliation{Departamento de F\'{\i}sica, Universidad de
Extremadura, E-06071 Badajoz, Spain}
\author{Mariano L\'opez de Haro}
\email{malopez@servidor.unam.mx}
\homepage{http://xml.cie.unam.mx/xml/tc/ft/mlh/} \affiliation{Centro
de Investigaci\'on en Energ\'{\i}a, Universidad Nacional Aut\'onoma
de M\'exico (UNAM), Temixco, Morelos 62580, M{e}xico}
%\thanks{}
\date{\today}
\begin{abstract}
The rational function approximation method, density functional
theory, and NVT Monte Carlo simulation are used to obtain the
density profiles of multicomponent hard-sphere mixtures near a
 planar hard wall. Binary mixtures with a size ratio 1:3 in
which both components occupy a similar volume are specifically
examined. The results indicate that the present version of density
functional theory yields an excellent overall performance. A
reasonably accurate behavior of the rational function approximation
method is also observed, except in the vicinity of the first
minimum, where it may  even predict unphysical negative values.
\end{abstract}


 \maketitle

%\narrowtext

\section{Introduction}
\label{sect_1} Many interesting physical phenomena such as wetting
and adsorption involve fluids and their mixtures at a solid-fluid
interface. It is therefore not surprising that a lot of work has
been devoted to these problems within the last three decades. The
methods that have been used to study them include the integral
equation approach of liquid state theory (see for instance Refs.\
\cite{HAB76, H78, PH84, PH85, HCD94, DAS97, ODHQ97, NHSC98}),
density functional approaches (see for instance Refs.\
\cite{TMvSG89, PH84, PG97, P99, RD00, ZR00, Z01, CG01}) and computer
simulation (see for instance Refs.\ \cite{SH78, TMvSG89, DH93,
NHSC98, RNG05}).

A rather simplified but essentially correct physical picture of
adsorption may be obtained if one considers the solid surface as a
planar smooth hard wall and the fluid as consisting of hard spheres
(HS). Interestingly enough, this simple model is capable of
accounting for the most important feature resulting from the
interaction between the fluid particles and the wall, namely the
strongly oscillatory nature of the  density distribution profile of
the fluid particles in the interfacial region. The particles are
depleted from the surface of the wall due to excluded volume effects
and this is the source of the density oscillations. A particular
realization of such a model is obtained from a binary HS mixture in
which one of the species is taken to have an infinite diameter and
to be in vanishing concentration. In this instance, the
wall-particle pair correlation functions lead immediately to the
density profiles. In fact, the availability of the analytical
solution of the Percus--Yevick (PY) equation for additive HS
mixtures obtained by Lebowitz \cite{L64} allowed Henderson \emph{et
al.} \cite{HAB76} to derive the density profile of a HS fluid near a
hard wall within the PY approximation already thirty years ago. More
recently, Noworyta \emph{et al.} \cite{NHSC98} also used the PY
theory to study two binary mixtures of HS near a hard wall. In the
same paper they also performed grand canonical ensemble simulations
for this system and used a version of density functional theory
(DFT) to derive the density profiles of both species. Their results
indicated that the second order PY theory provided in general the
best agreement with the simulation results and that the DFT was a
little less accurate. It is interesting to point out that for the
most asymmetric case that they examined, that of size ratio 1:3,
anomalies were observed, namely a huge discrepancy between theory
and simulation, for which they could find no explanation. This work
has served as a motivation for the present paper. Here we also
tackle the problem of binary HS mixtures at a planar hard wall. In
addition to the PY theory, we use the results of the rational
function approximation (RFA) method \cite{YSH98}, a different
version of DFT \cite{mfmt}, and NVT Monte Carlo simulation. Apart
from complementing the work of Ref.\ \cite{NHSC98}, our aim is to
assess the value of both the RFA method and of our version of DFT to
deal with this problem in a situation where both theories are
subject to rather stringent conditions, namely when there is a
disparate size of the diameters but both species occupy a similar
volume.

The paper is organized as follows. In order to make it
self-contained, in Sec.\ \ref{sect_2} we provide a brief summary of
the RFA method for the structural properties of a multicomponent HS
mixture and state the result for the planar hard wall limit in which
the concentration of one of the species goes to zero while its
diameter goes to infinity. The explicit derivation of such a limit
is made in the Appendix. Section \ref{sect_3} is devoted to the
outline of a recent version of DFT derived by one of us \cite{mfmt}.
This is followed in Sec.\ \ref{sect_4} by a description of the
simulation details. In Sec.\ \ref{sect_5} we present the results
derived with the two theoretical approaches as well as a comparison
with the simulation data. The paper is closed in Sec.\ \ref{sect_6}
with further discussion and some concluding remarks.

\section{The Rational Function Approximation Method}
\label{sect_2}

We start by considering an $(N+1)$-component fluid of (additive) HS
of diameters $\sigma_i$, mole fractions $x_i$ (with
$i=0,1,\ldots,N$), and total  bulk number density $\overline{\rho}$.
The partial  bulk number density of species $i$ is
$\overline{\rho}_i=x_i\overline{\rho}$ and the packing fraction of
the mixture is $\eta=(\pi/6)\overline{\rho}\langle \sigma^3\rangle$,
where
\beq
\langle \sigma^n\rangle=\sum_{i=0}^{N} x_i \sigma_i^n
\label{n1}
\eeq
denotes the $n$th moment of the size distribution.

According to the RFA for HS mixtures \cite{YSH98,HYS07}, the Laplace
transform
\beq
G_{ij}(s)=\int_{\sigma_{ij}}^\infty \dd r\, e^{-sr}r g_{ij}(r)
\label{m1}
\eeq
of $rg_{ij}(r)$, where $g_{ij}(r)$ is the radial distribution
function for the pair $ij$, is explicitly given by
\beq
\label{3.6}
G_{ij}(s)=\frac{e^{-\sigma_{ij} s}}{2\pi s^2} \left[{\sf
L}(s)\cdot {\sf B}^{-1}(s)\right]_{ij},
\eeq
where $\sigma_{ij}=(\sigma_i+\sigma_j)/2$ and ${\sf L}(s)$
and ${\sf B}(s)$ are $(N+1)\times (N+1)$ matrices given by
\beq
\label{3.7}
L_{ij}(s)=L_{ij}^\zero+L_{ij}^\one s+L_{ij}^\two s^2,
\eeq
\beq
{B}_{ij}(s)= (1+\alpha s)\delta_{ij}-{
 A}_{ij}(s),
\label{1}
\eeq
\beqa
\label{3.8}
A_{ij}(s)&=&\overline{\rho}_i\left[\varphi_2(\sigma_i s)\sigma_i^3
L_{ij}^\zero +\varphi_1(\sigma_i s)\sigma_{i}^2
L_{ij}^\one\right.\nn &&\left. +\varphi_0(\sigma_{i} s)\sigma_{i}
L_{ij}^\two\right].
\eeqa
In Eqs.\ \eqref{3.7}--(\ref{3.8}),
\beq
\label{2.9}
\varphi_n(x)\equiv x^{-(n+1)}\left(\sum_{m=0}^n \frac{(-x)^m}{m!}-
e^{-x}\right),
\eeq
\beq
\label{3.13}
L_{ij}^\zero=\lambda+\lambda'\sigma_j+2\lambda'\alpha-
\lambda\sum_{k=0}^N \overline{\rho}_k\sigma_k L_{kj}^\two,
\eeq
\beq
\label{3.14}
L_{ij}^\one=\lambda\sigma_{ij}+\frac{1}{2}\lambda'\sigma_i\sigma_j
+(\lambda+\lambda'\sigma_i)\alpha-\frac{1}{2}\lambda\sigma_i
\sum_{k=0}^N \overline{\rho}_k\sigma_k L_{kj}^\two, \eeq where \beq
\lambda\equiv \frac{2\pi}{1-\eta},\quad \lambda'\equiv
\frac{6\pi\eta}{(1-\eta)^2}\frac{\langle \sigma^2\rangle}{\langle
\sigma^3\rangle}.
\label{2}
\eeq

In principle, the matrix ${\sf L}^\two$ and the scalar $\alpha$ can
be chosen arbitrarily without violating any basic condition
\cite{YSH98,HYS07}. In particular, the choice $L_{ij}^\two=\alpha=0$
gives the PY solution \cite{L64,BH77}. One can go beyond this
approximation by prescribing given contact values
$g_{ij}(\sigma_{ij} )$ and the associated thermodynamically
consistent  (reduced) isothermal compressibility $\chi$, what fixes
${\sf L}^\two$ and $\alpha$ \cite{YSH98,HYS07}. Specifically,
\beq
\label{3.17}
{L_{ij}^\two}={2\pi\alpha\sigma_{ij}}g_{ij}(\sigma_{ij} ),
\eeq
while $\alpha$ is the smallest real root of an algebraic equation of
degree $2(N+1)$. Here, following the method of Ref.\ \cite{SYH05},
we will take for $g_{ij}(\sigma_{ij})$ the following extension to
mixtures of the Carnahan--Starling--Kolafa contact value \cite{K86}:
\beqa
g_{ij}(\sigma_{ij})&=&\frac{1}{1-\eta}+\frac{3 \eta}{2
\left(1-\eta\right)^2}\frac{\mt}{\mth}\frac{\sigma_i
\sigma_j}{\sigma_{ij}}+\frac{\eta^2(5-2\eta+2\eta^2)}{12(1-\eta)^3}\nn
&&\times\left(\frac{\mt}{\mth}\frac{\sigma_i
\sigma_j}{\sigma_{ij}}\right)^2+\frac{\eta^2(1+\eta)}{6(1-\eta)^2}\left(\frac{\mt}{\mth}\frac{\sigma_i
\sigma_j}{\sigma_{ij}}\right)^3.\nn
\label{eCSK3}
\eeqa
These contact values are thermodynamically consistent with
Boubl\'{\i}k's equation of state for HS mixtures \cite{boublik}. The
associated (reduced) isothermal compressibility is
\beqa
\chi&=&\left[\frac{1}{(1-\eta)^2}+\frac{6\eta}{(1-\eta)^3}\frac{\mo\mt}{\mth}\right.
\nn &&\left.+\eta^2\frac{27-8\eta-8\eta^2+4\eta^3}{3(1-\eta)^4}
\frac{\mt^3}{\mth^2}\right]^{-1}.
\label{n2}
\eeqa

Now we assume that the mole fraction of the spheres  of species
$i=0$ vanishes ($x_0\to 0$) and that their diameter is infinitely
larger than those of the other species ($\sigma_0/\sigma_i
\to\infty, i\geq 1$), in such a way that $x_0\sigma_0^n\ll \mn$ for
$n\leq 3$. In that case, species 0 does not contribute to either the
total packing fraction $\eta$  or the average values $\mo$, $\mt$,
and $\mth$, i.e,
\beq
\mn\to \sum_{i=1}^N x_i\sigma_i^n,\quad n\leq 3.
\label{n3}
\eeq
Under those conditions a particle of species $i=0$ is seen as a
planar hard wall by the $N$-component mixture made of particles of
species $i=1,2,\ldots,N$. By carefully taking the limits $x_0\to 0$
and $\sigma_0\to\infty$ (with the constraint $x_0\sigma_0^n\ll \mn$
for $n\leq 3$), it is proven in the Appendix that the \emph{local}
density of particles of species $i$ at a distance $z$ from the wall
is
\beq
\rho_i(z)=\overline{\rho}_i\gamma_i(z),
\label{m2}
\eeq
where the function $\gamma_i(z)$ is the inverse Laplace transform of
the function defined by Eqs.\ \eqref{46}, \eqref{m1b}--\eqref{47}.
As said in connection with Eqs.\ \eqref{3.6}--\eqref{3.8},
\eqref{3.13}, \eqref{3.14}, and \eqref{3.17},  the PY theory for
$\gamma_i(z)$ is reobtained by setting $\alpha=0$. Our extended RFA
theory is obtained by imposing Eqs.\ \eqref{eCSK3} and \eqref{n2}
and finding the corresponding value of $\alpha$. Note that, within
the RFA method, expressions for $g_{ij}$ and $\chi$ other than those
of Eqs.\ (\ref{eCSK3}) and \eqref{n2} could equally be used. The
choice we have made relies on the fact that these constitute well
tested and reliable approximations.

\section{Density functional theory}
\label{sect_3}

DFT is based on the property that, for a given interatomic
potential, the grand potential $\Omega$, or, equivalently, the
intrinsic Helmholtz free energy $F$, is a unique functional of the
one-body density profile $\rho_i(\rr)$ \cite{evans}. For an
$N$-component fluid mixture at given temperature $T$, total volume
$V$, chemical potential $\mu_i$, and external potential $V_i^{\rm
ext}(\rr)$ for each component, the equilibrium density profile
$\rho_i(\rr)$ minimizes the grand potential functional
 \bb
 \Omega\left[\{\rho_i\}\right]=F\left[\{\rho_i\}\right]+\sum_{i=1}^N\int\dd\rr\rho_i(\rr)[V_i^{\rm
 ext}(\rr)-\mu_i].
 \ee
The ideal contribution to the intrinsic free energy functional is
known exactly:
 \bb
 F_{\rm id}\left[\{\rho_i\}\right]=k_B
 T\sum_{i=1}^N\int\dd\rr\rho_i(\rr)\left\{\ln\left[\rho_i(\rr)\Lambda_i^3\right]-1\right\},
 \ee
where $k_B$ is the Boltzmann constant and $\Lambda_i$ is the thermal
de Broglie wavelength of species $i$. Thus, in the task of finding
the appropriate free energy functional one can focus only on the
excess part
 \bb
F_{\rm ex}\left[\{\rho_i\}\right]= F\left[\{\rho_i\}\right]-F_{\rm
id}\left[\{\rho_i\}\right].
 \ee
Once the expression for the intrinsic Helmholtz free energy
functional is known, the density distribution is  given explicitly
by
 \bb
 \rho_i(\rr)=\Lambda_i^{-3}\exp[c_i^{(1)}(\rr)+\beta\mu_i-\beta V_i^{\rm
 ext}(\rr)],
\label{Alex}
 \ee
 where $\beta \equiv 1/(k_B T)$ and
 $c_i^{(1)}(\rr)$ is the one-body direct correlation function,
 \bb
 c_i^{(1)}(\rr)=-\beta\frac{\delta F_{\rm
 ex}\left[\{\rho_j\}\right]}{\delta\rho_i(\rr)}.
 \ee

Because the knowledge of the free energy functional provides full
description of the model under study, practically any implementation
of DFT requires some explicit approximation for the functional
$F_{\rm ex}$. In the fundamental measure theory \cite{ros}, the
functional $F_{\rm ex}$ of a mixture of HS is assumed to take the
form
 \bb
 F_{\rm
 ex}\left[\{\rho_i\}\right]=k_BT\int\dd\rr\,\Phi(\{n_\alpha(\rr)\}),
\label{Fex}
 \ee
 where the excess free energy density,
  $\Phi(\{n_\alpha(\rr)\})$, depends only on the system averaged
 fundamental measures of the particles
 \bb
 n_\alpha(\rr)=\sum_{i=1}^N\int\dd\rr'\rho_i(\rr')\omega_i^{(\alpha)}(\rr-\rr').
 \ee
 The weight functions $\omega_i^{(\alpha)}(\rr)$ characterize the
 geometry of particles. The minimal space of the weight
functions is generated by the basis
 \bb
 \omega_i^{(3)}(\rr)=\Theta(\sigma_i/2-r),
 \label{o1}
 \ee
 \bb
 \omega_i^{(2)}(\rr)=\delta(r-\sigma_i/2),
 \ee
 \bb
 \mbox{\boldmath$\omega$}_i^{(2)}(\rr)=\frac{\rr}{r}\delta(r-\sigma_i/2),
 \label{o3}
 \ee
where $\Theta(r)$ is the Heaviside step function and $\delta(r)$ is
the Dirac delta function. The other weight functions are
proportional to those given by Eqs.~(\ref{o1})--(\ref{o3}), namely
$\omega_i^{(1)}(\rr)=\omega_i^{(2)}(\rr)/2\pi \sigma_i$,
$\omega_i^{(0)}(\rr)=\omega_i^{(2)}(\rr)/\pi \sigma_i^2$,
$\mbox{\boldmath$\omega$}_i^{(1)}(\rr)=\mbox{\boldmath$\omega$}_i^{(2)}(\rr)/2\pi
\sigma_i$.

In this work, we apply the following form of the free energy density
\cite{mfmt}:
 \beqa
 \Phi&=&-n_0\ln(1-n_3)+\frac{n_1n_2-{\bf n}_1\cdot{\bf
 n}_2}{1-n_3}+(n_2^3-3n_3{\bf n}_2\cdot{\bf n}_2)\nn
 &&\times
\frac{8(1-n_3)^2\ln(1-n_3)+8n_3-(15/2)n_3^2+2n_3^3}{108\pi
n_3^2(1-n_3)^2}.\nn
\label{Phi}
\eeqa
 In the limit of a bulk fluid, the vectorial weighted densities
${\bf n}_1$ and ${\bf n}_2$ vanish, $n_0=\overline{\rho}$,
$n_1=(\overline{\rho}/2)\langle \sigma \rangle$,
$n_2=\pi(\overline{\rho}/2)\langle \sigma^2 \rangle$, $n_3=\eta$,
and therefore  excess free energy density becomes equivalent to that
following from Boubl\'{\i}k's equation of state \cite{boublik}. In
the low density limit the functional given by Eqs.\ \eqref{Fex} and
\eqref{Phi} is equivalent to that originally proposed by Rosenfeld,
which underlies the PY compressibility  solution. It is well known,
however, that the PY compressibility  equation of state
overestimates the pressure at high densities. The functional
presented here includes corrections which more accurately
extrapolate from low to high density states \cite{mfmt}.

In the problem at hand, the external potential represents the
interaction with a planar hard wall, so that \bb V_i^{\rm
 ext}(\rr)=\left\{\begin{array}{ll}
\infty & {\rm for \; } z<\sigma_{i}/2 \\
0 & {\rm for \; } z>\sigma_{i}/2,
\end{array}\right.
\label{potext}
\ee
where $z$ is the coordinate normal to the
wall.

\section{Simulation details}
\label{sect_4}

A binary mixture of HS confined in a box of dimensions
$(L_x,L_y,L_z)$ was simulated using the NVT Monte Carlo method. The
particles were sorted out to cells and the linked-list method was
used. Periodic boundary conditions were applied in the $x$ and $y$
directions, whereas flat hard  walls were located at $z=0$ and
$z=L_z$. We chose $L_x=L_y=10$ and $L_z=30$ in units of $\sigma_{1}$
(the diameter of the small spheres) for all the simulations.

The initial parameters of the simulations correspond to the average
densities and the average concentrations. The connection between
average and bulk values is not initially known so that the bulk
values are evaluated in the region of the simulation box $z\sim
L_z/2$, where no influence of  the walls is expected. It is worth
mentioning that the bulk values can be alternatively determined
using two separate boxes connected by the same chemical potentials
(one with and the other without hard walls) or using the Gibbs
ensemble. However, we have found the simple NVT method (which does
not involve particle insertions) to be more efficient and less
problematic, especially for asymmetric mixtures and/or for high
densities.

There are two common procedures as to how an initial configuration
can be prepared. The random shooting method starting with the
species of the largest diameter fails to converge at high packing
fractions. On the other hand, the box can be initially filled up
with particles placed at the crystal configuration, usually an fcc
lattice. The disadvantage of the latter is the need of ``relaxing''
the system, a process that may take too long and may even turn to be
untractable. We present a novel method which suffers none of the
above mentioned problems and can be easily generalized to the model
of an arbitrary number of components. First of all, both ``large''
and ``small' particles are inserted randomly into the box with the
only restriction of no overlapping of the like particles. Then the
system is left to propagate keeping the interaction between the like
particles hard, whereas the penetrability of the unlike particles is
restricted by the potential \bb u_{ij}(r)=\left\{\begin{array}{ll}
\epsilon \left[1+a(\sigma_{ij}-r)/\sigma_1\right] & {\rm for \; } r<\sigma_{ij} \\
0 & {\rm for \; } r>\sigma_{ij},
\end{array}\right.
\label{eq:pot}
 \ee
 where the value $a=1$ of the constant turned
up to be convenient. Choosing the initial reduced temperature
(typically $T^*\equiv k_B T/\epsilon=0.05$), the system was
gradually cooled down. We found this algorithm to be extremely
efficient in creating a non-overlapping configuration (typically
within one minute using a standard PC).

The system is subsequently equilibrated over more than $10^7$ MC
steps  and the bulk densities and density profiles are averaged
over another $4\times10^8$ moves. The average runs were divided
into $20$ subaverages to estimate the standard deviations.

\section{Results}
\label{sect_5}
\begin{figure*}[t]
\includegraphics[width=2 \columnwidth]{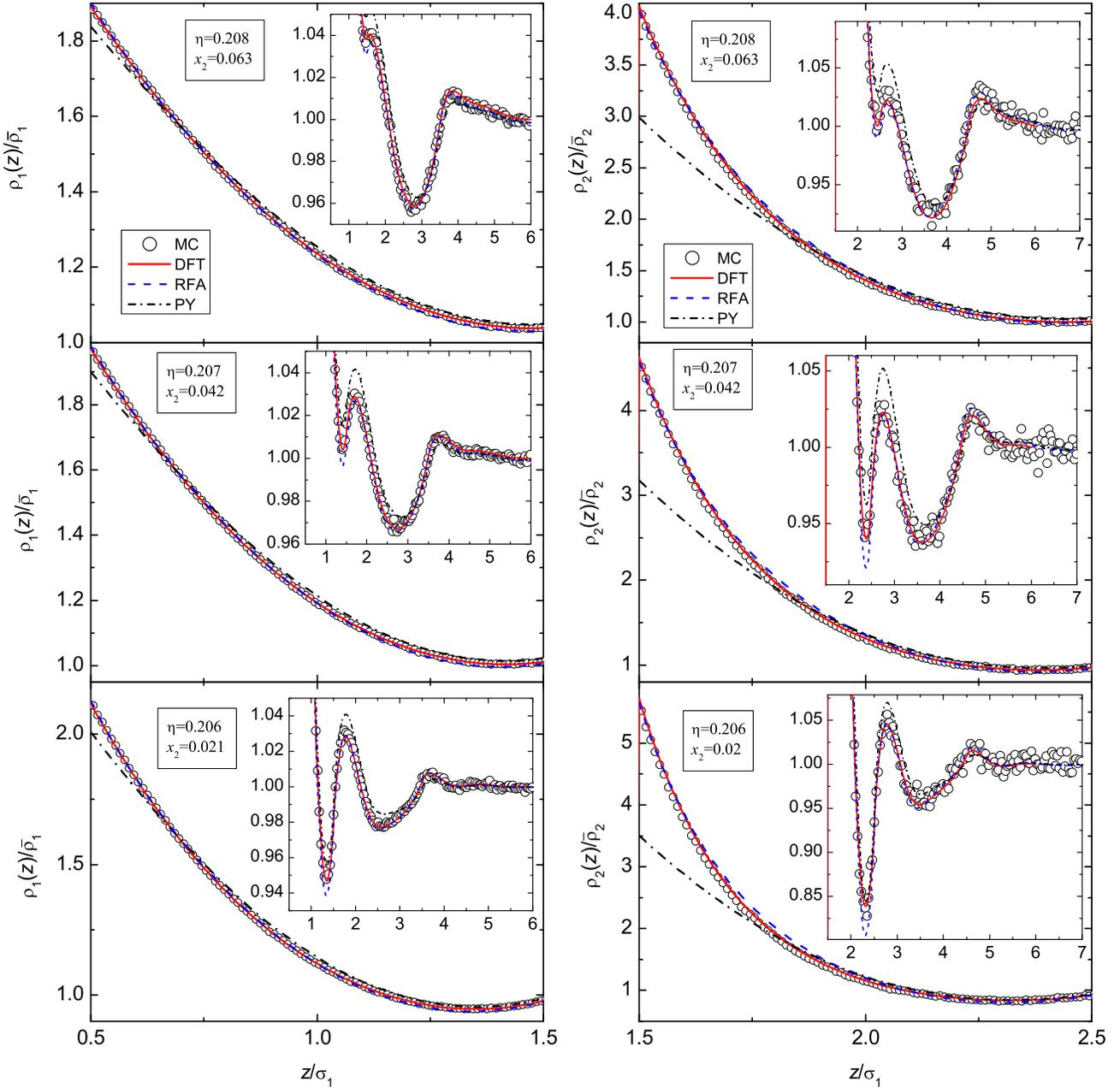}
\caption{{(Color online) Density profiles close to the wall for both
species in the binary HS mixtures with $\sigma_2/\sigma_1=3$ having
a total packing fraction close to $\eta=0.2$. Top panels:
$\eta=0.208$ and $x_2= 0.063$; middle panels:  $\eta=0.207$ and
$x_2=0.042$; bottom panels:  $\eta=0.206$ and $x_2= 0.021$. Solid
lines: DFT; dashed lines: RFA results, dash-dotted lines: PY theory.
The symbols are the results of simulation. The insets provide a
wider range and more details of the resulting structure.}
\label{fig1}}
\end{figure*}
\begin{figure*}[t]
\includegraphics[width=2\columnwidth]{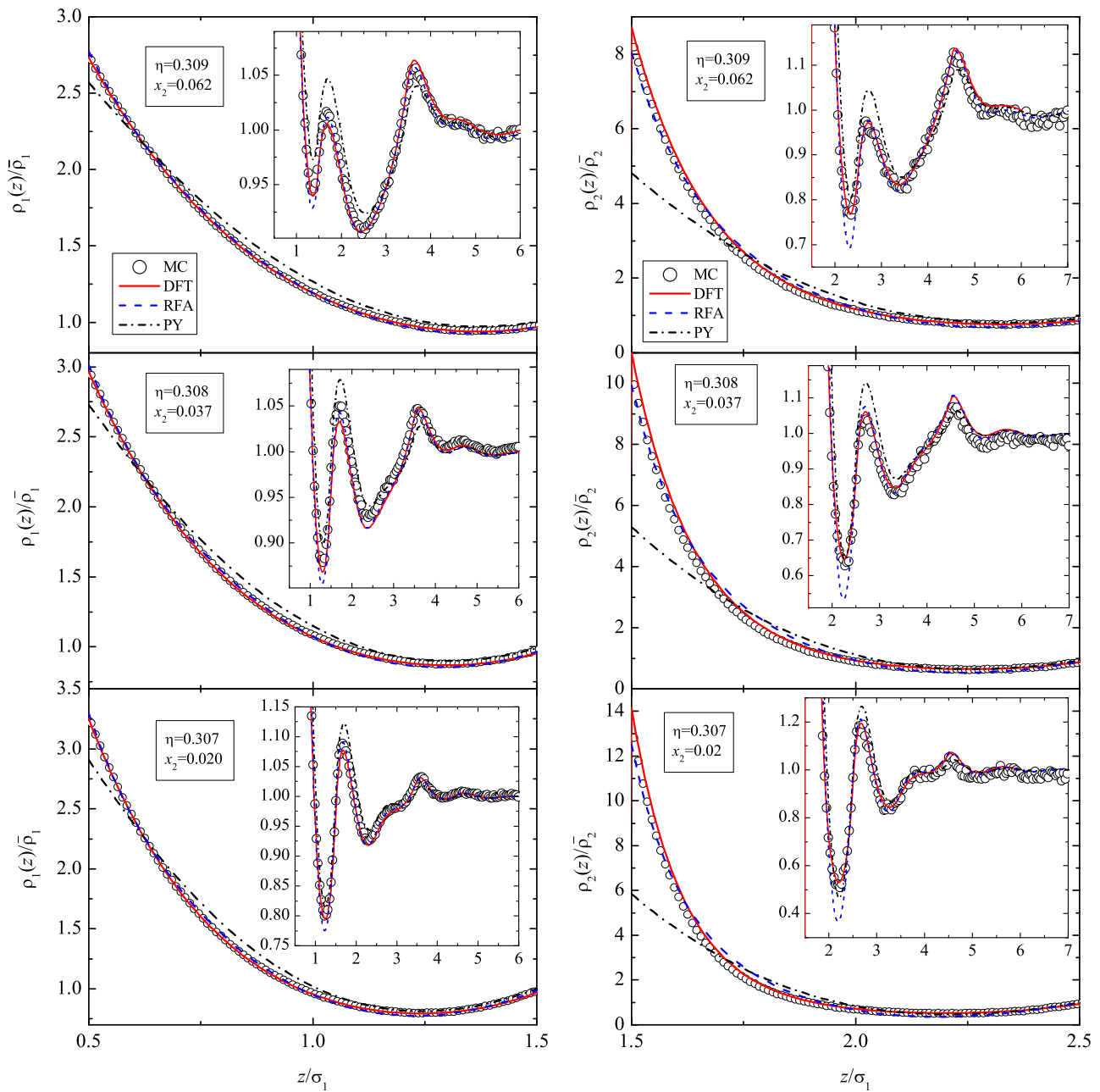}
\caption{{(Color online) The same as Fig.\ \protect\ref{fig1} but
with a total packing fraction close to $\eta=0.3$. Top panels:
$\eta=0.309$ and $x_2= 0.062$; middle panels:  $\eta=0.308$ and
$x_2=0.037$; bottom panels:  $\eta=0.307$ and $x_2= 0.020$.}
\label{fig2}}
\end{figure*}
\begin{figure*}[t]
\includegraphics[width=2 \columnwidth]{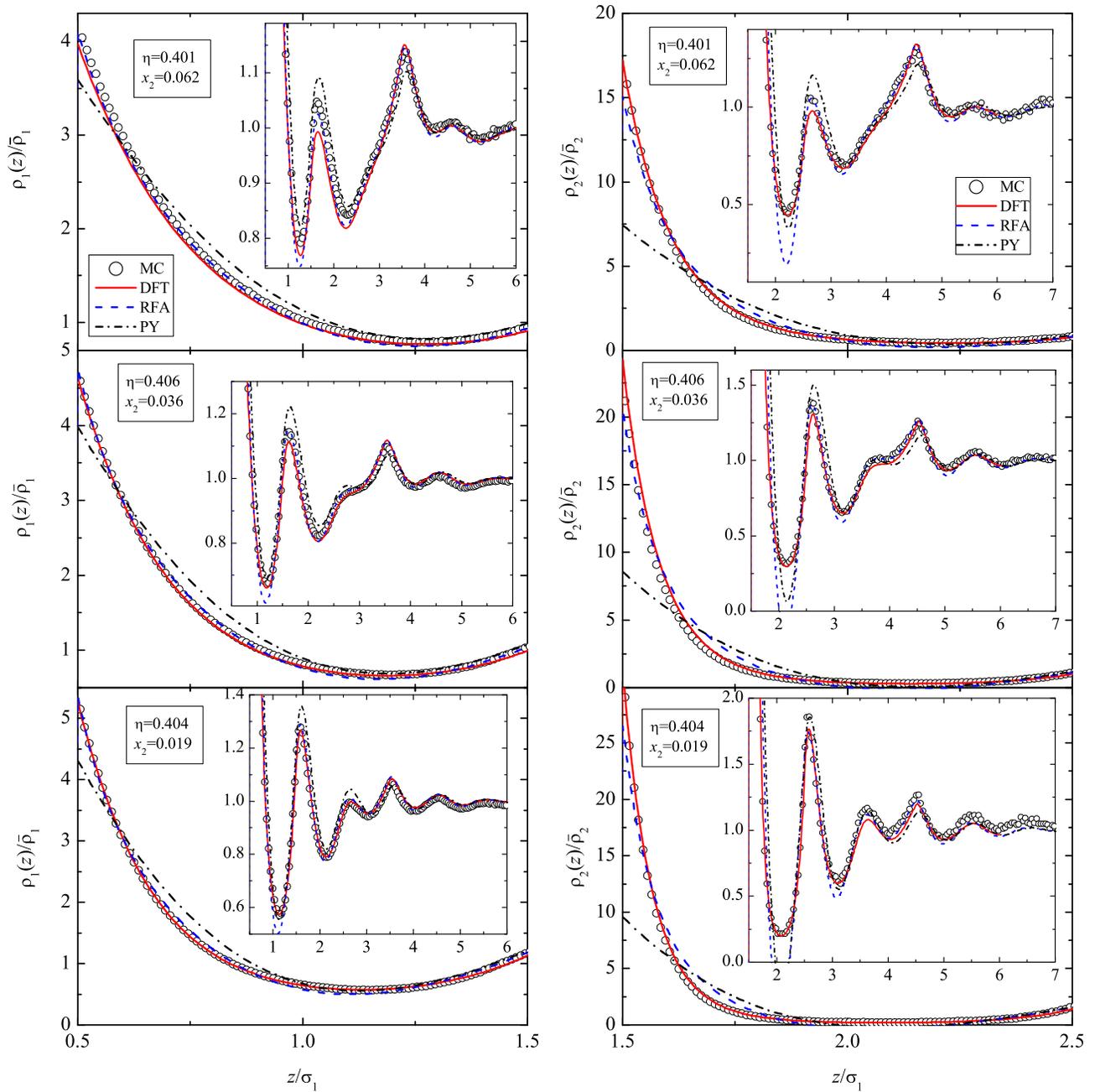}
\caption{{(Color online) The same as Fig.\ \protect\ref{fig1}   but
with a total packing fraction close to $\eta=0.4$. Top panels: $x_2=
0.062$ and $\eta=0.401$; middle panels: $x_2=0.036$ and
$\eta=0.406$; bottom panels: $x_2= 0.019$ and $\eta=0.404$.}
\label{fig3}}
\end{figure*}
\begin{figure*}[t]
\includegraphics[width=2 \columnwidth]{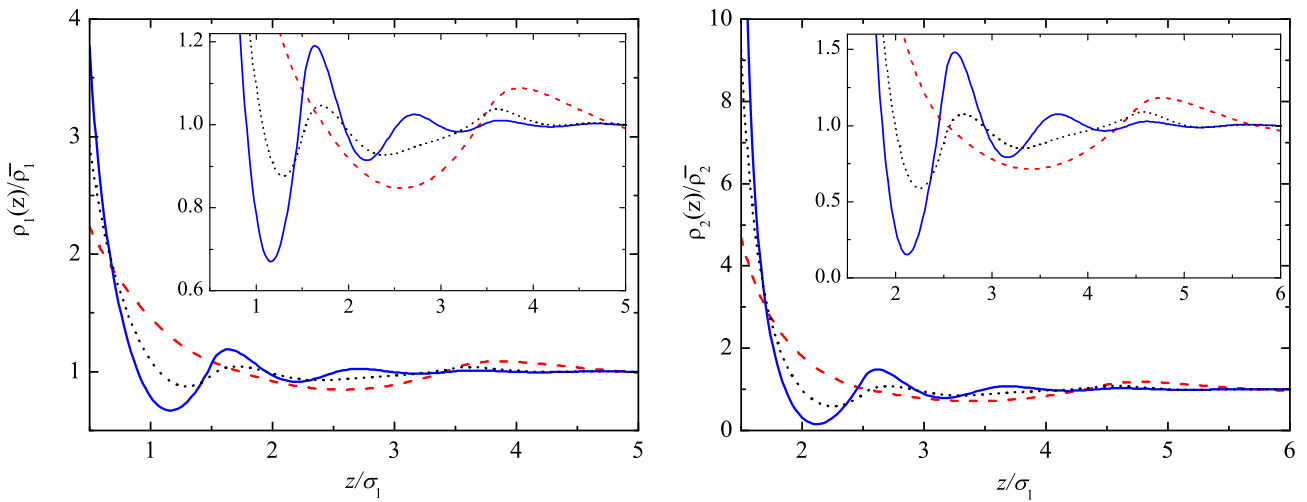}
\caption{{(Color online) Density profiles close to the wall for both
species in the binary HS mixtures with $\sigma_2/\sigma_1=3$ having
a total packing fraction $\eta=0.3$, as predicted by the RFA. Dashed
lines: $x_2=\frac{10}{37}\simeq 0.27$ ($\eta_1/\eta_2=0.1$); dotted
lines: $x_2=\frac{1}{28}\simeq 0.036$ ($\eta_1/\eta_2=1$); solid
lines: $x_2=\frac{1}{271}\simeq 0.0037$ ($\eta_1/\eta_2=10$). The
insets provide a more detailed view of the oscillations.}
\label{fig4}}
\end{figure*}
\begin{figure}[t]
\includegraphics[width=\columnwidth]{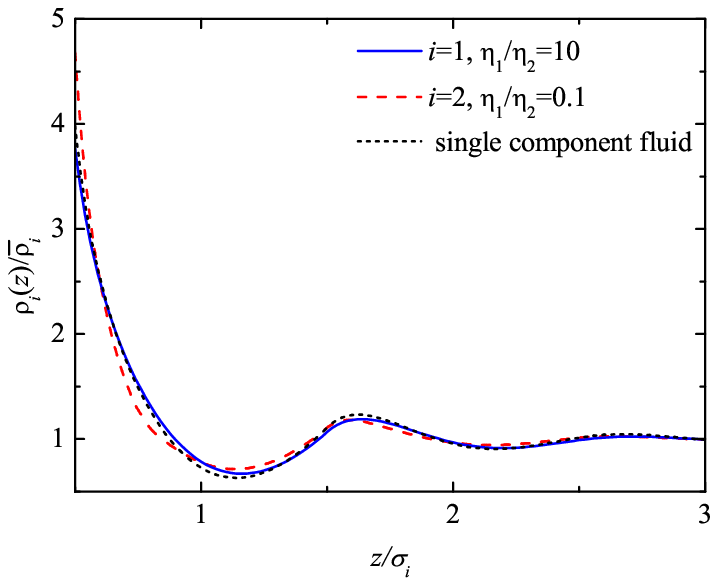}
\caption{{(Color online) Density profiles close to the wall  in the
binary HS mixtures with $\sigma_2/\sigma_1=3$ having a total packing
fraction $\eta=0.3$, as predicted by the RFA. Solid line:
$\rho_1(z)/\bar{\rho}_1$  versus $z/\sigma_1$ for
$x_2=\frac{1}{271}\simeq 0.0037$ ($\eta_1/\eta_2=10$); dashed line:
$\rho_2(z)/\bar{\rho}_2$  versus $z/\sigma_2$ for
$x_2=\frac{10}{37}\simeq 0.27$ ($\eta_1/\eta_2=0.1$). The dotted
line corresponds to the single component case.}
\label{fig5}}
\end{figure}

In this Section we report the results we have obtained with the
previous approaches, namely the RFA method, DFT, and simulation, for
the density profiles of binary HS mixtures in the presence of a
planar hard wall. In order to test the theories under extreme
conditions, we consider cases in which the sizes of the two
components in the mixture are disparate but both occupy a similar
volume.

For the sake of illustration, in Figs.\ \ref{fig1}--\ref{fig3} we
present the density profiles for both components of nine binary
mixtures with a common size ratio $\sigma_2/\sigma_1=3$ and a total
(bulk) packing fraction $\eta\simeq 0.2$ (Fig.\ \ref{fig1}),
$\eta\simeq 0.3$ (Fig.\ \ref{fig2}), and $\eta\simeq 0.4$ (Fig.\
\ref{fig3}). For each packing fraction, three different compositions
have been considered:   $x_2\simeq 0.06$, so that
$\eta_2/\eta_1\simeq 1.7$ (top panels);  $x_2\simeq 0.04$, so that
$\eta_2/\eta_1\simeq 1.1$ (middle panels); and $x_2\simeq 0.02$, so
that $\eta_2/\eta_1\simeq 0.55$ (bottom panels). Here,
$\eta_i=(\pi/6)\overline{\rho}_i\sigma_i^3$ ($i=1,2$) denotes the
partial (bulk) packing fraction of species $i$.  Since $\eta$ and
$x_2$ are measured in the bulk, they present minor deviations with
respect to the imposed average values in each case. Apart from the
simulation data and our theoretical approaches, the figures also
include the PY results.

Without any doubt, the DFT is the one that produces the best overall
agreement. As far as the RFA method is concerned, one finds that it
also does a very good job in general, being particularly accurate
not only at the contact value (which is of course an input in this
approach), but also caters very well for the second maximum and the
rest of the oscillations. These nice features worsen in the vicinity
of the first minimum, particularly for the density profile of the
bigger species at $\eta\simeq 0.4$, where the RFA method may lead to
the unphysical prediction of a negative value. As expected, the PY
theory yields poorer contact values and gets worse as the total
packing fraction increases, yielding a negative value for the first
minimum  of $\rho_2(z)$ at $\eta\simeq 0.4$ and $x_2\simeq 0.02$. In
any case, it indeed accounts for the oscillatory character of the
profiles.

Figures \ref{fig1}--\ref{fig3} show that a rich and complex
structure of the local densities $\rho_1(z)$ and $\rho_2(z)$ appears
for all the cases considered. This is related to the fact that
$\eta_1\sim\eta_2$ and so both species compete for the available
volume. As the total density increases so does the nonuniformity of
the density profiles, as reflected by the contrast between the
values at contact and at the first minimum, especially in the case
of the large spheres. Moreover, as the mole fraction $x_2$ of the
large spheres decreases, the characteristic wavelength of the
oscillations decreases. To  better understand this phenomenon,
imagine a mixture with a mole fraction $x_2$ such that
$\eta_1\ll\eta_2$. In that case, the large spheres (species 2) are
practically unaffected by the presence of the small ones, so the
local density $\rho_2(z)$ is the same as that of a single component
fluid at the same packing fraction and oscillates with a
characteristic wavelength of the order of $\sigma_2$. As for the
small spheres (species 1), their density profile is dominated by the
presence of the large spheres and so the wavelength of $\rho_1(z)$
is also of the order of $\sigma_2$. In the other extreme situation,
namely when $\eta_2\ll\eta_1$, we have the opposite situation: the
small spheres behave as a single component fluid and enslave the
density profile of the large spheres, so that both $\rho_1(z)$ and
$\rho_2(z)$ oscillate with a characteristic wavelength of the order
of $\sigma_1$. The interplay between both extreme cases occurs when
$\eta_1\sim\eta_2$, giving rise to the superposition of both length
scales, as observed in Figs.\ \ref{fig1}--\ref{fig3}. To illustrate
the transition from $\eta_1\ll\eta_2$ to $\eta_2\ll\eta_1$, in Fig.\
\ref{fig4} we plot the density profiles predicted by the RFA for
mixtures with $\sigma_2/\sigma_1=3$, $\eta=0.3$, and
$x_2=\frac{10}{37}\simeq 0.27$ (which corresponds to
$\eta_1/\eta_2=0.1$), $x_2=\frac{1}{28}\simeq 0.036$ (which
corresponds to $\eta_1/\eta_2=1$), and $x_2=\frac{1}{271}\simeq
0.0037$ (which corresponds to $\eta_1/\eta_2=10$). A ratio 1:10 in
the partial packing fractions is enough to make the species with the
largest packing fraction to behave almost as a single component
fluid. In fact, Fig.\ \ref{fig5} shows that the curve of
$\rho_2(z)/\overline{\rho}_2$ in the case $x_2=\frac{10}{37}\simeq
0.27$ ($\eta_1/\eta_2=0.1$) and the curve of
$\rho_1(z)/\overline{\rho}_1$ in the case $x_2=\frac{1}{271}\simeq
0.0037$  ($\eta_1/\eta_2=10$) look like very similar when each one
is plotted as a function of the corresponding scaled distance
$z/\sigma_i$. It must be noticed that, as observed in Fig.\
\ref{fig5},  the case $\eta_1/\eta_2=10$ is actually closer to the
single component fluid than the case $\eta_1/\eta_2=0.1$. In the
latter case one has $x_1/x_2=2.7$, $x_1\sigma_1/x_2\sigma_2=0.9$,
and $x_1\sigma_1^2/x_2\sigma_2^2=0.3$, so that the influence of the
species 1 is not entirely negligible. On the other hand, the ratios
are $x_2/x_1=0.0037$, $x_2\sigma_2/x_1\sigma_1=0.011$, and
$x_2\sigma_2^2/x_1\sigma_1^2=0.033$ in the case $\eta_1/\eta_2=10$.

\section{Concluding Remarks}
\label{sect_6}

The results presented in the preceding Section deserve some further
discussion. To begin with, inspired by the work of Noworyta \emph{et
al.} \cite{NHSC98}, we have revisited the problem of determining the
structure of binary HS mixtures in the presence of a  planar hard
wall using three of the different approaches that have been proposed
in the literature to deal with it. Concerning the RFA method, its
main advantage is that of allowing for a completely analytical
description, as it also occurs with the PY theory, avoiding at the
same time the thermodynamic inconsistency problem present in the
latter. It is fair to say that, under the rather extreme conditions
that we tested it, this method is a reasonable compromise between
accuracy and simplicity. Whether the difficulties associated with
the prediction of a negative first minimum of $\rho_2(z)$ at
$\eta\simeq 0.4$ and $x_2\lesssim 0.04$ could be avoided by the
introduction of additional parameters in the method remains to be
assessed. It should be mentioned in passing that for mixtures with
$\eta\simeq 0.3$  this unphysical behavior is not predicted, even
for $x_2\to 0$. Moreover, in the case of pure ternary mixtures, the
agreement between the results for the structural properties of the
RFA method and simulation has been shown to be highly satisfactory
\cite{MMYSH02}. Finally, although we have illustrated the results
for the case of binary mixtures, a further asset of the development
we have presented applies in principle to any multicomponent mixture
of HS near a  planar hard wall. On the other hand, the excellent
performance of the DFT in the approximation introduced in Ref.\
\cite{mfmt} has been already pointed out. However the merit of the
proposed excess free energy functional cannot be overlooked and
might be useful for other purposes as well. As a final point, it
should be stressed that the simulation method reported in this paper
is also novel and allowed us to obtain results for situations that
were problematic ten years ago. In particular, it allowed us to
confirm the assertion made by Roth and Dietrich \cite{RD00}
concerning the non existence of the anomaly reported by Noworyta
\emph{et al.} \cite{NHSC98} for the mixture with size ratio 1:3. Our
hope and expectation is that this method proves useful as well for
other interesting systems, some of which we plan to examine in the
future.

Before closing this paper and for the sake of setting a wider
perspective for the results we have presented, a few further
comments are in order. A simple and yet realistic model of colloidal
dispersions consists of a highly asymmetric binary hard-sphere
mixture in which the large spheres stand for the colloidal particles
and the small spheres represent solvent or polymer molecules.
Therefore, our results may also be applied in these systems, which
are nowadays easily amenable for experimental examination, and open
up the possibility of investigating solvation forces and other
interesting physical phenomena associated with particles of
different sizes competing for interfacial positions in the presence
of walls. We plan to pursue some of these issues in the future. On
the other hand, the possible extension of our work to deal with
non-hard particles  remains to be explored. In the case of the  RFA
method, structural properties of sticky hard-sphere mixtures and
other systems have already been derived \cite{HYS07}.  As for the
DFT approach, modifications would  be required. Thus it seems that
much work remains to be done before such extensions become a
reality. Again, we plan to explore these and other possibilities in
future studies.

\acknowledgments

The research of Al.M. has been
 partially supported by the Ministry of Education, Youth, and Sports of
the Czech Republic under Project No.\ LC 512 (Center for
Biomolecules and Complex Molecular Systems) and by the Grant Agency
of the Czech Republic under Projects No.\ 203/06/P432 and No.\
203/05/0725. The research of S.B.Y. and A.S. has been supported by
the Ministerio de Educaci\'on y Ciencia (Spain) through Grant No.\
FIS2004-01399 (partially financed by FEDER funds) and by the Junta
de Extremadura-Consejer\'{\i}a de Infraestructuras y Desarrollo
Tecnol\'ogico. M.L.H. acknowledges the financial support of
DGAPA-UNAM through Project IN-110406. Two of the authors (S.B.Y. and
A.S.) are grateful to the Centro de Investigaci\'on en Energ\'{\i}a
(UNAM, Mexico) for its hospitality during a two-week visit in
January-February 2007, when this work was finished.

\appendix*

\section{The wall limit in the RFA\label{appA}}

The limit $x_0\to 0$ implies that the  row $i=0$ of the $(N+1)\times
(N+1)$ matrix  ${\sf A}$ defined by Eq.\ \eqref{3.8} vanishes, so
that
\beq
A_{ij}=\widetilde{A}_{ij}(1-\delta_{i0})(1-\delta_{j0})+A_{i0}(1-\delta_{i0})\delta_{j0}.
\label{4}
\eeq
Here, $\widetilde{A}_{ij}$ is a \textit{projected} $N\times N$
matrix with $i,j \geq 1$. In fact, $\widetilde{A}_{ij}$ is the
matrix corresponding to an $N$-component mixture in the absence of
species $0$. Inserting (\ref{4}) into Eq.\ (\ref{1}), we have
\beq
B_{ij}=\widetilde{B}_{ij}(1-\delta_{i0})(1-\delta_{j0})+(1+\alpha
s)\delta_{i0}\delta_{j0}-A_{i0}(1-\delta_{i0})\delta_{j0},
\label{5}
\eeq
where $\widetilde{B}_{ij}$ is the $N\times N$ matrix
\beq
\widetilde{ B}_{ij}= (1+\alpha s)\delta_{ij}-\widetilde{
 A}_{ij}, \quad i,j\geq 1.
\label{6}
\eeq
It can be checked that the $(N+1)\times (N+1)$ inverse matrix ${\sf
B}^{-1}$ is
\beqa
\left({\sf B}^{-1}\right)_{ij}&=&\left(\widetilde{\sf
B}^{-1}\right)_{ij}(1-\delta_{i0})(1-\delta_{j0})+ (1+\alpha
s)^{-1}\delta_{i0}\delta_{j0}\nn &&-C_{i}(1-\delta_{i0})\delta_{j0},
\label{7}
\eeqa
where $\widetilde{\sf B}^{-1}$ is the $N\times N$ inverse matrix of
$\widetilde{\sf B}$ and the elements $C_i$ are
\beq
C_i=\frac{1}{1+\alpha s}\sum_{j=1}^N \left(\widetilde{\sf
B}^{-1}\right)_{ij}A_{j0}.
\label{10}
\eeq
Insertion of Eq.\ (\ref{7}) into Eq.\ (\ref{3.6}) yields
\beqa
2\pi s^2 e^{\sigma_{ij}s}G_{ij}&=& \sum_{k=1}^N
L_{ik}\left(\widetilde{\sf
B}^{-1}\right)_{kj}(1-\delta_{j0})+\frac{L_{i0}}{1+\alpha
s}\delta_{j0}\nn &&+ \sum_{k=1}^N L_{ik}C_k \delta_{j0}.
\label{11}
\eeqa
In particular, if $i,j \geq 1$,
\beq
2\pi s^2 e^{\sigma_{ij}s}G_{ij}= \sum_{k=1}^N L_{ik}\left({\sf
B}^{-1}\right)_{kj},\quad i,j\geq 1,
\label{12}
\eeq
where  we have taken into account that $\left(\widetilde{\sf
B}^{-1}\right)_{ij}=\left({\sf B}^{-1}\right)_{ij}$ if $i,j\geq 1$.
Equation \eqref{11}  implies that, as expected, the $N$-component
mixture is not affected by the presence of the species 0 in the
infinite dilution limit $x_0\to 0$. On the other hand, setting
$i\geq 1$ and $j=0$ in Eq.\ (\ref{11}), we have
\beq
2\pi s^2 e^{\sigma_{i0}s}G_{i0}= \frac{L_{i0}}{1+\alpha
s}+\frac{2\pi s^2}{1+\alpha s} \sum_{j=1}^N
e^{\sigma_{ij}s}G_{ij}A_{j0}, \quad i\geq 1,
\label{13}
\eeq
where use has been made of Eqs.\ (\ref{10}) and (\ref{12}). The
cross function  $G_{i0}$ (with $i=1,\ldots,N$) is related to the
spatial correlation between the diluted species $0$ and the species
$i$ of the true $N$-component mixture. We see from Eq.\ \eqref{13}
that $G_{i0}$ is expressed in terms of the matrix $G_{ij}$ of the
$N$-component mixture and the cross elements $L_{i0}$ and $A_{j0}$.

 Let us now introduce
the \textit{shifted} radial distribution function
\beq
\gamma_{i}(z)=g_{i0}(z+\sigma_{0}/2),
\label{38}
\eeq
where $z\geq \sigma_i/2$ represents  the distance from the center of
a sphere of species $i$ to the \emph{surface} of a sphere of species
$0$. In Laplace space,
\beq
G_{i0}(s)=e^{-\sigma_{0}s/2}\left[\frac{\sigma_{0}}{2}\N_{i}(s)-\N'_{i}(s)\right],
\label{39}
\eeq
where
\beq
\N_{i}(s)=\int_{\sigma_i/2}^\infty \dd z\, e^{-s z}\gamma_{i}(z)
\label{40}
\eeq
is the Laplace transform of $\gamma_{i}(z)$  and $\N'_{i}(s)=\dd
\N_{i}(s)/\dd s$.

Thus far, the diameter $\sigma_0$ is arbitrary as long as Eq.\
\eqref{n3} is satisfied. Now we take the wall limit
($\sigma_0\to\infty$). In that case, the function $\gamma_i(z)$ has
a clear meaning as the ratio between the local density of particles
of species $i$ at a distance $z$ from the wall, $\rho_i(z)$, and the
corresponding density in the bulk,
$\overline{\rho}_i=\rho_i(\infty)$. In the wall limit $\N_i'(s)$ can
be neglected versus $\sigma_{0}\N_i(s)$ in Eq.\ \eqref{39}.
Therefore,
\beqa
\N_{i}(s)&=&{2}\lim_{\sigma_0\to\infty}{\sigma_0}^{-1}
e^{\sigma_{0}s/2}G_{i0}(s)\nn &=& e^{-\sigma_{i}s/2}\left[
\frac{\mathcal{L}_{i0}(s)}{\pi s^2(1+\alpha s)}+\frac{2}{1+\alpha
s}\right.\nn &&\times \left.\sum_{j=1}^N
e^{\sigma_{ij}s}G_{ij}(s)\mathcal{A}_{j0}(s)\right], \quad i\geq 1,
\label{46}
\eeqa
where in the last step use has been made of Eq.\ (\ref{13}) and
\beq
\mathcal{L}_{i0}(s)\equiv\lim_{\sigma_0\to
\infty}\sigma_0^{-1}L_{i0}(s),\quad
\mathcal{A}_{i0}(s)\equiv\lim_{\sigma_0\to
\infty}\sigma_0^{-1}A_{i0}(s) .
\label{46bis}
\eeq
From Eqs.\ \eqref{3.7}, (\ref{3.8}), (\ref{3.13}), (\ref{3.14}), and
(\ref{3.17}) one gets
\beq
\mathcal{L}_{i0}(s)=\mathcal{L}_{i0}^\zero+ \mathcal{L}_{i0}^\one s+
\mathcal{L}_{i0}^\two s^2,
\label{m1b}
\eeq
\beq
\mathcal{L}_{i0}^\zero=\lambda'-\pi \alpha
\lambda\sum_{j=1}^N\overline{\rho}_j\sigma_j g_{j0}(\sigma_{j0}),
\label{41}
\eeq
\beq
\mathcal{L}_{i0}^\one=\frac{\lambda}{2}+\frac{\lambda'}{2}\sigma_i-\pi
\alpha \frac{\lambda}{2}\sigma_i
\sum_{j=1}^N\overline{\rho}_j\sigma_j g_{j0}(\sigma_{j0}),
\label{42}
\eeq
\beq
\mathcal{L}_{i0}^\two=\pi \alpha g_{i0}(\sigma_{i0}),
\label{43}
\eeq
\beqa
\mathcal{A}_{i0}(s)&=&\overline{\rho}_i\left[\varphi_2(\sigma_i
s)\sigma_i^3 \mathcal{L}_{i0}^\zero +\varphi_1(\sigma_i
s)\sigma_{i}^2 \mathcal{L}_{i0}^\one \right.\nn
&&\left.+\varphi_0(\sigma_{i} s)\sigma_{i}
\mathcal{L}_{i0}^\two\right].
\label{47}
\eeqa
It must be noted that the parameter $\alpha$ appearing in Eqs.\
\eqref{46} and \eqref{41}--\eqref{43} is independently obtained from
the RFA solution for the $N$-component mixture. Regarding the
contact values $g_{i0}(\sigma_{i0})$, they are obtained by taking
the limit $\sigma_0\to\infty$ in Eq.\ \eqref{eCSK3}, namely
\beqa
g_{i0}(\sigma_{i0})&=&\frac{1}{1-\eta}+\frac{3 \eta}{
\left(1-\eta\right)^2}\frac{\mt}{\mth}{\sigma_i}+\frac{\eta^2(5-2\eta+2\eta^2)}{3(1-\eta)^3}\nn
&&\times\left(\frac{\mt}{\mth}{\sigma_i}\right)^2+\frac{4\eta^2(1+\eta)}{3(1-\eta)^2}\left(\frac{\mt}{\mth}{\sigma_i}\right)^3.\nn
\label{eCSK3w}
\eeqa

Taking into account that $G_{ij}(s)=s^{-2}+{\cal O}(s^0)$ for small
$s$, it is possible to prove that
\beq
\lim_{s\to 0}s\Gamma_{i}(s)=1,
\label{new1}
\eeq
which implies the physical condition
\beq
\lim_{z\to\infty}\gamma_{i}(z)=1.
\label{new2}
\eeq

\end{document}